\documentclass[12pt]{article}
\usepackage{graphicx}
\usepackage{amsmath}
\usepackage{amssymb}
\usepackage{hyperref}
\numberwithin{equation}{section}

\def\bC{\mathbb{C}}

\def\bR{\mathbb{R}}

\def\cC{\mathcal{C}}
\def\cF{\mathcal{F}}
\def\cH{\mathcal{H}}
\def\cM{\mathcal{M}}
\def\cN{\mathcal{N}}
\def\cT{\mathcal{T}}
\def\fg{{\displaystyle\mathfrak{g}}}
\def\fS{{\displaystyle\mathfrak{S}}}
\def\Bo{\mathrm{Bo}}
\def\HS{\mathrm{HS}}
\def\HK{\mathrm{HK}}
\def\SL{\mathrm{SL}}
\def\SU{\mathrm{SU}}
\def\SO{\mathrm{SO}}

\def\Aut{\mathrm{Aut}}

\def\Obj{\mathrm{Obj}}
\def\Hom{\mathrm{Hom}}

\def\rank{\mathop{\mathrm{rank}}}

\def\sl{\mathfrak{sl}}

\def\inc#1{\vcenter{\hbox{\includegraphics[scale=.2]{#1}}}}

\begin{document}

\begin{titlepage}

\vbox{}
\vskip 3cm
\begin{center}
\def\thefootnote{\fnsymbol{footnote}}
{\Large \bfseries
On 2d TQFTs whose values are\\[.1em]
 holomorphic symplectic varieties
}

\vskip 1.2cm
Gregory W. Moore$^1$ and Yuji Tachikawa$^2$\footnote[1]{on leave from IPMU}
\vskip 0.4cm
$^1$ NHETC and Department of Physics and Astronomy, Rutgers University,\\
 Piscataway, NJ 08855–0849, USA

\bigskip

$^2$ School of Natural Sciences, Institute for Advanced Study, \\
Princeton, New Jersey 08504, USA

\vskip 1.5cm

\textbf{abstract}
\end{center}
{For simple and simply-connected complex algebraic group ${G_{\bC}}$, we conjecture the existence of a functor $\eta_{G_{\bC}}$ from the category of 2-bordisms to the category of holomorphic symplectic varieties with Hamiltonian action, such that gluing of boundaries  corresponds to the holomorphic symplectic quotient with respect to the diagonal action of ${G_{\bC}}$.
We describe various properties of $\eta_{G_{\bC}}$ obtained via string-theoretic analysis.
Mathematicians are urged to construct $\eta_{G_{\bC}}$ rigorously.
}

\bigskip

\begin{center}
\itshape To appear in the proceedings of String-Math 2011
\end{center}
\end{titlepage}

\section{Introduction}

Physicists have been gathering evidence for the existence of a class of interacting superconformal quantum field theories
in six dimensions with $\cN=(2,0)$ supersymmetry. These theories are believed to have an ADE classification, i.e.~they are classified by a connected, simply connected and simply laced compact Lie group $G$, and
 appear to lie at the heart of a large number of results and constructions in
physical mathematics. We will  denote these theories by $S[G]$.
For a discussion of 6d $\cN=(2,0)$ theory aimed at mathematicians,
see \cite{Witten:2009at,Witten:2009mh}.

Let $C$ be a punctured Riemannian surface and consider the theories $S[G]$ on 
 $\mathbb{R}^{1,3} \times C$. Combining the conformality of $S[G]$ with a 
partial topological twist one can argue that the resulting theory 
depends only on the conformal structure of $C$. Consequently, these partially 
twisted theories  should satisfy factorization properties reminiscent of those of two-dimensional
topological and conformal field theories. In Section 2 below we provide a slightly more precise
description of these factorization properties, although we hasten to add that
this description will not be completely satisfactory
to mathematicians.  This state of affairs can, however, be ameliorated.
Several geometric objects, such as certain branches of moduli spaces
of vacua, can be associated to the theories $S[G]$  on $C$, and these geometric objects inherit the factorization
properties of the parent theory. The assignment of such geometric objects to
the data of $C$ \emph{is} something which is susceptible to rigorous mathematical
discussion. In this note we provide a simple example.
Namely, we show  how  the ``maximal-dimension
Higgs branch of the theory $S[G]$ on $C$\,''  provides an example of
a two-dimensional topological field theory valued in a symmetric monoidal category which is
not simply a  category of vector spaces. This formulation  captures some results originally found
and described in physics language in the literature.
See, for examples, \cite{Argyres:2007cn,Argyres:2007tq,Gaiotto:2008sa,Gaiotto:2008ak,Gaiotto:2008nz,
Gaiotto:2009gz,Gaiotto:2009we,Benini:2009gi,Nanopoulos:2009xe,Benini:2009mz,Benini:2010uu,Chacaltana:2010ks,Tachikawa:2010vg,Hanany:2010qu}.

The existence of this two dimensional field theory relies
on the existence of certain holomorphic symplectic manifolds satisfying properties
listed in Section 3.  We hope mathematicians find the formulation of this topological
field theory sufficiently precise and interesting to provide a rigorous construction of the required
manifolds.

\section{Physical Background}

In this section we sketch in a little more detail the sense in which the theories $S[G]$
define a generalization of two-dimensional conformal field theory. Readers who are
only interested in the two-dimensional topological field theory
or those who prefer rigorously formulated mathematics can safely skip this
section and proceed to Section 3.

Quantum field theories can have ``defects,'' or ``defect operators.'' These are operators
or degrees of freedom which can be placed on positive codimension subspaces of spacetime.
Thus, for example,  a local operator is a point defect, a Wilson line operator is a line defect, and so on.
The six-dimensional theories $S[G]$ have certain supersymmetric codimension-two defects.
For our purposes all we need to know about these defects is that they are specified by
 a homomorphism $\rho: \mathfrak{su}(2)\to \fg$.
  Let us now consider an
 oriented surface $C$ of genus $g$ with $n$ punctures with a Riemannian metric of finite volume.
 We   study the theory $S[G]$ on $\mathbb{R}^{1,3}\times C$ with the codimension-two
defects located at the punctures. We denote the punctures $p_a$ assigned with the homomorphism $\rho_a$ collectively as $D$, and refer to the pair $(C,D)$  as a decorated surface.

The theory  $S[G]$ on $\mathbb{R}^{1,3}\times C$ admits a partial topological twist
such that, when one takes the long distance limit, (i.e. the ``Kaluza-Klein reduction'')
the result is a  four-dimensional theory with
$d=4, \cN=2$ supersymmetry, which we will denote   $S_G(C, D)$. This construction
was introduced in  \cite{Klemm:1996bj,Witten:1997sc,Gaiotto:2009hg,Gaiotto:2009we}. These theories,
known as \emph{theories of class $S$},  have many 
properties. One is that the defects $D_a$ associated to punctures $p_a \in C$
have a global symmetry group \footnote{The physics literature is rather imprecise about which compact form  of the group one should choose, and the issue is nontrivial for reasons discussed in  \cite{Witten:2009at}. We will gloss over that point in this Section, and it will not  affect the more precise mathematical statements of Sections 3 and 4. }
   $G_{a}\subset G$, the commutant of the image of $\rho$.
 The global symmetry
group of the theory $S_G(C, D)$ includes $\prod_a G_{a} $.
Moreover, the theory $S_G(C,D)$   only
depends on the \emph{conformal structure} on $C$  (but up to some subtleties).
This property results from the  topological twisting.
Furthermore, the  space of coupling constants is the moduli space $\cM_{g,n}$ of complex structures on $C$,
and the Deligne-Mumford boundaries correspond to weak coupling limits of the theory.

In \cite{Gaiotto:2009we},  D. Gaiotto argued that at the boundaries of $\cM_{g,n}$
 the  theories $S_G(C, D)$   should exhibit  factorization properties analogous to
those enjoyed by two-dimensional conformal field theories and two-dimensional
topological field theories. Let us consider two
decorated surfaces  $(C_L , D_L)$ and $(C_R,D_R)$ and let us
pick   punctures $p_L \in C_L$ and $p_R \in C_R$ both with $\rho_{p_L}=\rho_{p_R}=0$ so that $G_{p_L}=G_{p_R}=G$.
Now, on the one hand, we can consider the product theory, and gauge the diagonal subgroup of its
global $G \times G$ symmetry.
In $d=4,\cN=2$ gauge theory the only essential parameter introduced in this operation is
a coupling constant $q=e^{2\pi i \tau}$.
Call the resulting theory $S_L \times_{G,q} S_R$.
On the other hand, choosing local coordinates $z_L, z_R$ at $p_L, p_R$ such that $p_{L,R}$ is at $z_{L,R}=0$,
we can form the
punctured surface $C_L \times_q C_R$ by setting $z_Lz_R=q$ with data $D_{LR}$ at the remaining punctures.
Gaiotto's crucial statement is that the two constructions lead to the same theory:
\begin{equation}\label{eq:gaiotto-glue}
S_L \times_{G,q} S_R = S_G( C_L \times_q C_R, D_{LR} ) .
\end{equation}

We caution the reader that for some pairs $(C,D)$ (e.g. when $C$ is a sphere with two punctures)
the theory $S_G(C,D)$
does not exist as a genuine four-dimensional theory, and hence in some situations
\eqref{eq:gaiotto-glue} must be
interpreted with care. Nevertheless, this  result points to the existence of
a  generalization of two-dimensional conformal field theory where
we have  a two-dimensional field theory whose target category is
something like the ``symmetric monoidal category of four-dimensional $\cN=2$ theories.''

The idea of a topological field theory whose target category is
a general symmetric monoidal category has appeared quite often before
in the formal study of topological field theories. One recent discussion may be
found in \cite{Kapustin:2010ta}. If $\cF$ is an $n$-dimensional
topological field theory then for any fixed compact $k$-dimensional manifold $K$
the functor $\cF_K$ whose domain is the the $(n-k)$-dimensional bordism category
and which is defined by   $\cF_K: M \mapsto \cF(M\times K)$
 is  an $(n-k)$ dimensional
topological field theory. This is a TFT version of Kaluza-Klein reduction
 along $K$. Therefore, $K \mapsto \cF_K$ is an
example of a $k$-dimensional field theory whose target category is the
category of  $(n-k)$-dimensional field theories.
In an analogous way, we would like to regard $S_G$ as a ``functor''
from decorated surfaces:   $S_{G}:  (C,D) \to  S_G(C,D)$,
where $C$ is endowed with a conformal structure, as an
example of a ``two-dimensional conformal field theory whose target category is
the category of  four-dimensional
$\cN=2$ theories.'' This statement can surely be made more precise, and it would
be worthwhile doing so. However, in the absence of a mathematically rigorous
formulation of an $\cN=2$, $d=4$ quantum field theory it cannot be made fully rigorous.

In any case, 
we will view    $S_{G}$ as something analogous to a functor.
Now,  physicists know that many different
 mathematical objects can be associated to a 4d $\cN=2$ theory ``functorially.''
Composing them with $S_G$, we expect to have  operations  which associate to $(C,D)$ simpler mathematical objects,
which might be rigorously formulated. Let us mention a few examples:
\begin{enumerate}
\item
Take the maximal-dimension Higgs branch $\cH(\cT)$ of a 4d $\cN=2$ theory $\cT$.
Denoting this operation by $\cH$, we consider the composed operation $\eta_G=\cH\circ S_G$.
This operation $\eta_G$ associates a hyperk\"ahler manifold to a punctured 2d surface $C$.
The construction preserves $\SU(2)$ R-symmetry, which is manifested as the $\SO(3)$ isometry of the hyperk\"ahler manifold rotating three complex structures. Supersymmetry implies the Higgs branch should be independent of coupling constant,
and hence it is  believed that $\eta_G(C)$ only depends on the topology of $C$.
Since $S_G$ is a `functor' and $\cH$ is a natural operation, we may expect that
$\eta_G$ is susceptible of a precise mathematical definition as
 a well-defined functor from the category of 2-bordisms to the category of hyperk\"ahler manifolds.
Due to a subtlety which we come back to in Section 5, we need to regard the image of $\eta_G$
as a holomorphic symplectic variety to define a genuine TQFT.
The $\SO(3)$ isometry of the hyperk\"ahler manifold is now manifested as the $\bC^\times$ action $\psi_t$ on the variety
under which the symplectic form $\omega$ is rescaled, i.e.~$\psi_t^*(\omega)=t^{-2}\omega$.
In the next section we will show that
 $\eta_G$ determines a 2d TQFT whose values are holomorphic symplectic varieties. In other words, $\eta_G(C)$ is the ``amplitude''  associated to the surface $C$.

\item We could also talk about  the composition $\cC\circ S_G$, where $\cC(\cT)$ of a 4d $\cN=2$ theory $\cT$ is
the ``Coulomb branch of $\cT$ on $\bR^3\times S^1$.'' As a holomorphic symplectic variety this is the algebraic integrable system canonically associated to $\cT$. (See, for examples \cite{Itoyama:1995nv,Donagi:1997sr} for reviews of early work and \cite{Gaiotto:2009hg} for a recent discussion.)  For a punctured Riemann surface $C$,
 $\cC\circ S_G(C)$ is the moduli space of a Hitchin system on $C$. It is to be emphasized that $\eta_G(C)$ is \emph{not} to be confused with $\cC \circ S_G(C)$. Rather, $S_G(C)$ is the ``3d mirror'' to $\cC\circ S_{G}(C)$
  \cite{Intriligator:1996ex,Gaiotto:2008ak}. Understanding the factorization properties of $\cC\circ S_G$ expected from physics appears to be challenging.

\item It should also be emphasized that there are many other objects which can be extracted from a ``4d $\cN=2$ theory.''  Another  class of objects is defined for each four-manifold $X$, and is called the partition function on $X$ and denoted by $Z_X$. For compact
 $X$ this will be a complex number. (In practice, $X$ is endowed with extra data such as parameters of equivariant
 cohomology, when it has symmetries, or the data of external gauge fields coupling to global symmetries, in which case
 it is a function of these parameters.)   $Z_X$ has factorization properties analogous to the correlation functions
 of a conformal field theory or of a topological field theory on $C$. There are some notable examples of this in
 the literature.
 For example $Z_{S^4}\circ S_G(C)$ for $G=\mathrm{SU}(2)$ turns out to be related to the correlation  functions of local
 operators on $C$ of the  Liouville conformal field theory \cite{Alday:2009aq}. This statement has been generalized
 to the 2d Toda field theory of type $G$ with $W$-symmetry in    \cite{Wyllard:2009hg}.
 The local operators are inserted at the punctures of $C$, and determined by the defect data.
 On the other hand, $Z_{S^3\times S^1} \circ S_G(C)$ is  independent of the complex structure.
It is in general a two-parameter deformation of the 2d $q$-deformed
  two-dimensional Yang-Mills theory in the zero-area limit \cite{Gadde:2009kb,Gadde:2010te,Gadde:2011ik}.

\end{enumerate}

The rest of the note is devoted to the formulation of $\eta_G=\cH\circ S_G$.
Conjecturally, the functor $\eta_G$ also exists for non-simply-laced $G$.
There is no known 6d $\cN=(2,0)$ theory corresponding to non-simply-laced Lie algebras.
Therefore, when $G$ is not simply-laced, one first puts a suitable
6d $\cN=(2,0)$ theory on $S^1$ with an automorphism twist to produce
5D super Yang-Mills theories with  gauge group $G^\vee$ \cite{Vafa:1997mh,Argyres:2006qr}. Compactifying such a theory on
the punctured surface $C$ produces a 3d theory, whose  moduli space of vacua contains one branch of the
form $\eta_G[C]$ and another branch which  is the moduli space of
a $G^\vee$-Hitchin system.

\section{Axioms}\label{sec:Axioms}
\subsection{The source and the target categories}
Here we describe the basic properties of our functor. First we specify the source and the target categories.
We take the source category to be the bordism category $\Bo_2$, i.e.~the objects are closed oriented
one-dimensional manifolds (i.e.~disjoint unions of multiple $S^1$s) and a morphism from $B_1$ to $B_2$ is a
two-dimensional oriented manifold $C$ whose boundary is $B_1 \sqcup (-B_2)$.
$\Bo_2$ is a symmetric monoidal category with duality under the standard operations.

The target category $\HS$ is a category of holomorphic symplectic varieties with Hamiltonian action defined as follows.
Let us start with the category structure:
\begin{itemize}
\item The elements of $\Obj(\HS)$ are complex algebraic semi-simple groups (including the trivial group $1$).
\item For ${G_{\bC}},{G_{\bC}}'\in \Obj(\HS)$, an element of $\Hom({G_{\bC}},{G_{\bC}}')$ is a triple $([X],{G_{\bC}},{G_{\bC}}')$.
Here $X$ is a holomorphic symplectic variety with a $\bC^\times$ action $\psi_t$ such that $\psi_t^*(\omega) = t^{-2} \omega$ where $t\in \mathbb{C}^\times$,
together with a holomorphic Hamiltonian action of ${G_{\bC}}\times {G_{\bC}}'$ commuting with $\psi_t$.
We identify an $X$ with symplectic form
$\omega$ with  an  $ X'$ with symplectic form $\omega'$ if there is a holomorphic isomorphism
$f: X \to X'$ commuting with the ${G_{\bC}} \times {G_{\bC}}'$ action and the $\bC^\times$ action
such that $f^*(\omega')=\omega$.
$[X]$ denotes the resulting equivalence class. To lighten the notation we write informally
$X \in \Hom({G_{\bC}},{G_{\bC}}')$ but one must bear in mind that a morphism has an ordered pair of groups ${G_{\bC}}$ and ${G_{\bC}}'$.
\item For $X\in\Hom({G_{\bC}}',{G_{\bC}})$ and $Y\in\Hom({G_{\bC}},{G_{\bC}}'')$, their composition $Y\circ X\in \Hom({G_{\bC}}',{G_{\bC}}'')$ is defined as the holomorphic symplectic quotient \begin{equation}
Y\circ X := X\times Y  // {G_{\bC}}= \{(x,y)\in X\times Y \ | \ \mu_X(x)+\mu_Y(y)=0 \} / {G_{\bC}}
\end{equation} where $\mu_X:X\to \fg_\bC^*$  and $\mu_Y:Y\to \fg_\bC^*$ are moment maps of the ${G_{\bC}}$ action on $X$ and $Y$, respectively.
Here $\fg_\bC$ is the Lie algebra of ${G_{\bC}}$.
\item The identity $id_{G_{\bC}}\in \Hom({G_{\bC}},{G_{\bC}})$ is $T^*{G_{\bC}}$ which has Hamiltonian ${G_{\bC}}\times {G_{\bC}}$ action.
\end{itemize}

We must check that these definitions define a category. The associativity of the composition
follows readily from the definition. To see that $T^*{G_{\bC}}$ acts as the identity consider,
for example $T^*{G_{\bC}}\circ X$.  We can identify $T^*{G_{\bC}} \cong {G_{\bC}} \times \fg_\bC$ using left- or right-invariant forms.
Denoting an element of $T^*{G_{\bC}}$ by $(g,a)$ the moment map condition is $a + \mu(x)=0$, which
eliminates $a$. The induced two-form on the solution space is  $G_\bC$-invariant  and  basic.
The quotient by ${G_{\bC}}$ allows us to gauge $g$ to $1$,   thus giving a holomorphic
isomorphism with the original space $X$  with its symplectic form.

The category $\HS$ is a symmetric monoidal category.
The monoidal structure is given by the following operations, which are obviously symmetric:
\begin{itemize}
\item For ${G_{\bC}},{G_{\bC}}'\in \Obj(\HS)$, ${G_{\bC}}\times {G_{\bC}}'\in \Obj(\HS)$ is the Cartesian product of groups.
\item  For $X\in \Hom({G_{\bC}},{G_{\bC}}')$ and $Y\in \Hom(H_\bC,H_\bC')$, $X\times Y\in \Hom({G_{\bC}}\times H_\bC,{G_{\bC}}'\times H_\bC')$ is also the Cartesian product of $X$ and $Y$.
\end{itemize}
Note that the trivial group $1$ is the unit under this operation.

The category $\HS$ also comes with duality.
Here by a symmetric monoidal category with duality we mean one such that \begin{itemize}
\item For any object $A$, there is a dual object $A^*$.
\item There are basic morphisms $p_A\in\Hom(A\times A^*, 1)$ and $q_A\in\Hom(1,A^*\times A)$ so that the following identities (sometimes called the ``zig-zag identities'' or ``S diagram'') hold:\begin{equation}
( p_A\times id_{A})\circ(id_{A}\times q_A) = id_{A}
\quad
(id_{A^*} \times p_A)\circ(q_A\times id_{A^*}) = id_{A^*},
.
\end{equation}
\end{itemize}
The duality structure for $\Bo_2$ is well known: It is simply the statement
that the standard S-cobordism is equal to the tube. The duality structure
for $\HS$ is given by:
\begin{itemize}
\item For an object ${G_{\bC}}\in\Obj(\HS)$, we define its dual to be ${G_{\bC}}$ itself.
\item We let $q_{G_{\bC}}\in \Hom(1,{G_{\bC}}\times {G_{\bC}})$ and $p_{G_{\bC}}\in \Hom({G_{\bC}}\times {G_{\bC}},1)$   both be $T^*{G_{\bC}}$; they trivially satisfy the zig-zag identities.
\end{itemize}

\subsection{Functor}\label{bar}

\label{defining}
Choose a simple, simply-connected algebraic group ${G_{\bC}}$.
We want to define a functor \begin{equation}
\eta_{G_{\bC}} : \Bo_2 \to \HS
\end{equation}  between symmetric monoidal categories with duality.
The image of the object $S^1$ is ${G_{\bC}}$: \begin{equation}
\eta_{G_{\bC}}(S^1)={G_{\bC}}.
\end{equation}
Next, because our  categories have duality, it suffices to specify the image of the basic corbordisms with one, two and three incoming circles:
\begin{alignat}{2}
U&=\inc{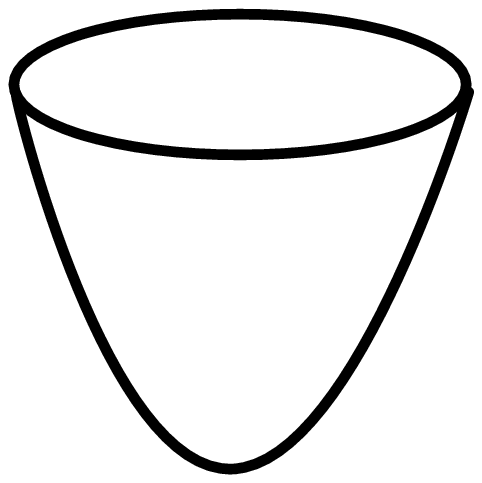}
&&\in\Hom(S^1,\varnothing), \\
V&=\inc{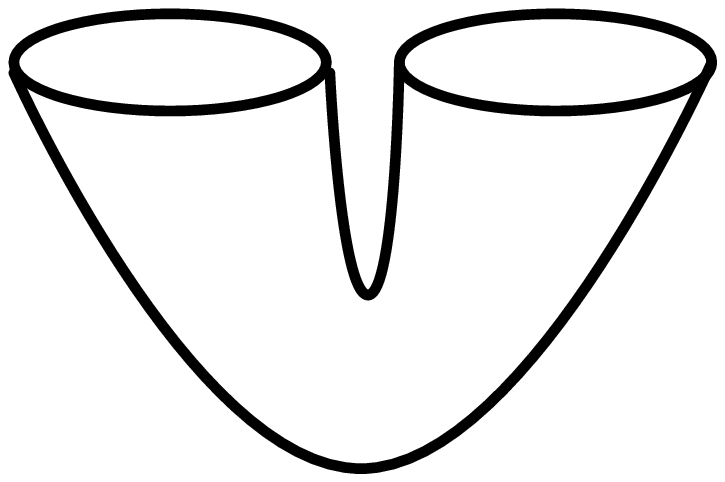}
&&\in\Hom(S^1\sqcup S^1,\varnothing),\\
W&=\inc{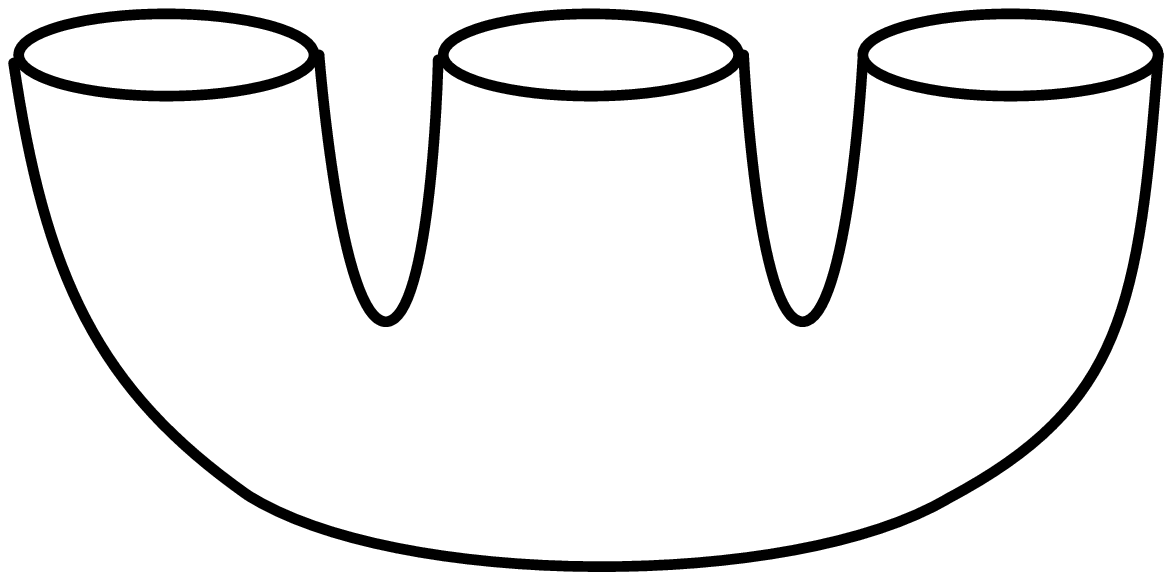}
&&\in \Hom(S^1 \sqcup  S^1 \sqcup S^1,\varnothing).
\end{alignat}
Our convention is that in all the figures all boundary components are incoming. Other morphisms
are then easily obtained by using $p_{S^1}$ and $q_{S^1}$.

Let us define varieties $U_{G_{\bC}}:=\eta_{G_{\bC}}(U)$ and $W_{G_{\bC}}:=\eta_{G_{\bC}}(W)$.
In order for $\eta_{G_{\bC}}$ to be a functor,  we must have $\eta_{G_{\bC}}(V)=T^*{G_{\bC}}$, and moreover, $U_{G_{\bC}}$ and $W_{G_{\bC}}$ have to satisfy  basic  sewing axioms of two-dimensional TFT (see e.g.~\cite{Moore:2006dw}). In our case these axioms
translate into the following statements:
\begin{itemize}
\item (Capping)  $U_{G_{\bC}} \circ W_{G_{\bC}} =T^*{G_{\bC}}$. This comes from the diagram: \begin{equation}
\eta_{G_{\bC}}(\inc{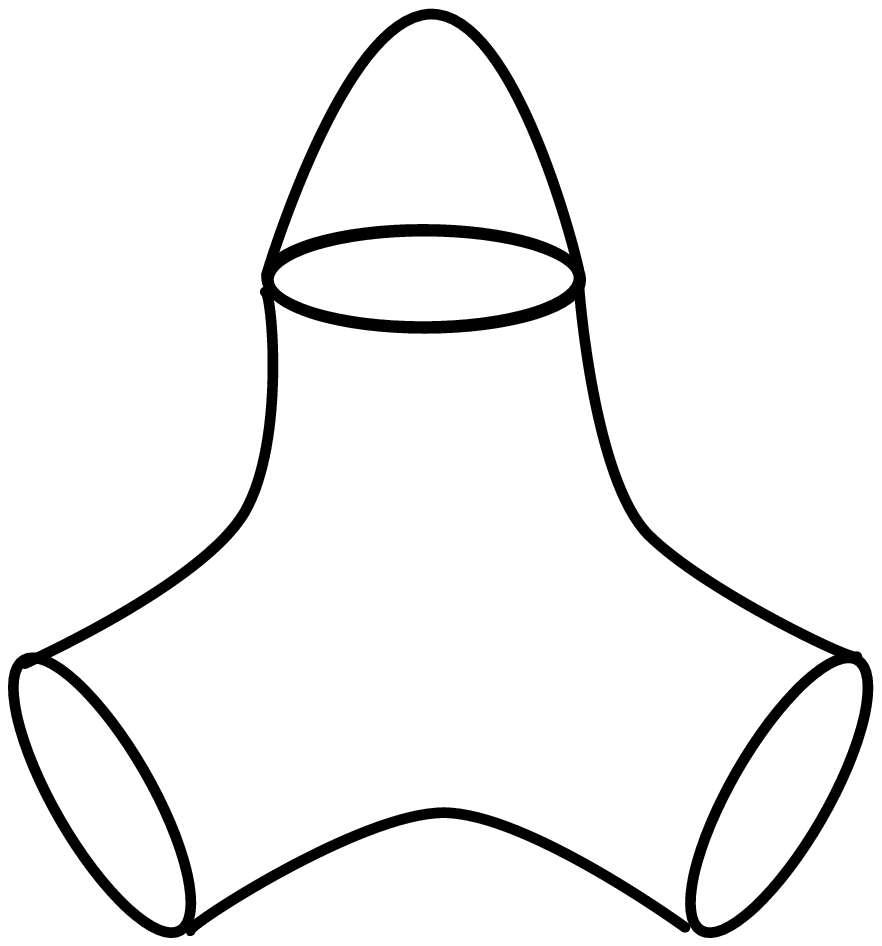})=\eta_{G_{\bC}}(\inc{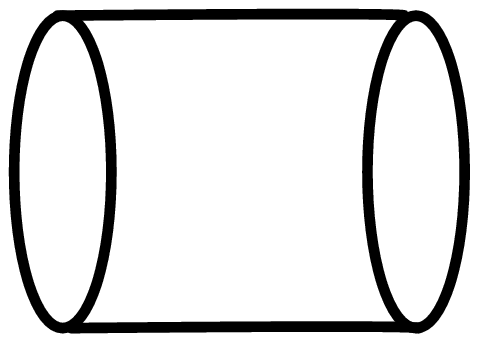}).
\end{equation}
\item (Commutativity) $W_{G_{\bC}}$ has an action of $\fS_3$ permuting the Hamiltonian ${G_{\bC}}^3$ action, i.e.~it has a holomorphic symplectic action of $\fS_3\ltimes {G_{\bC}}^3$. This comes from the diffeomorphism exchanging three boundaries of the pair of pants.
\item (Associativity) $W_{G_{\bC}} \circ W_{G_{\bC}}$ has an action of $\fS_4$ permuting the Hamiltonian ${G_{\bC}}^4$ action, i.e.~it has a holomorphic symplectic action of $\fS_4\ltimes {G_{\bC}}^4$. This comes from the following diagram:
\begin{equation}
\eta_{G_{\bC}}(\inc{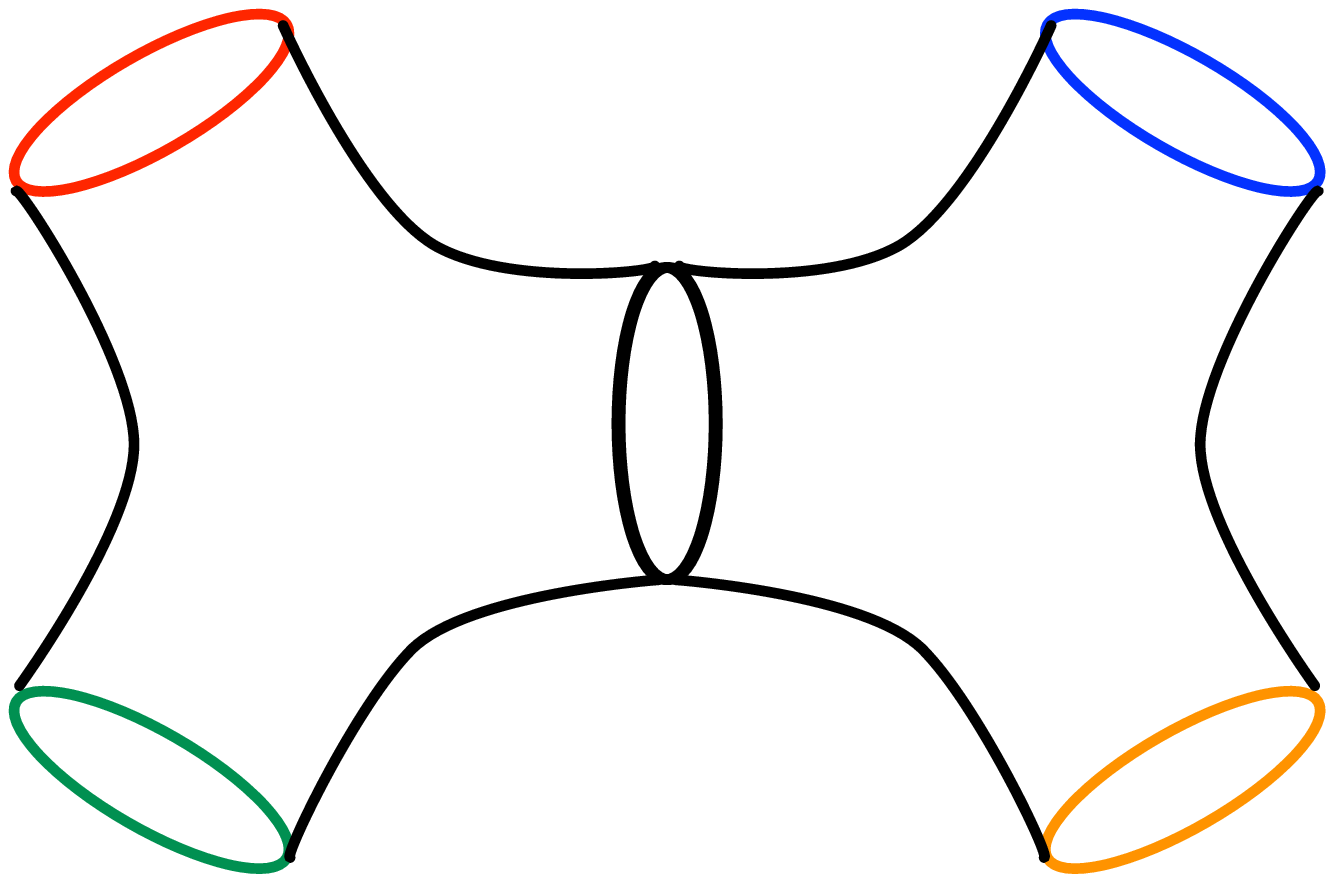})=\eta_{G_{\bC}}(\inc{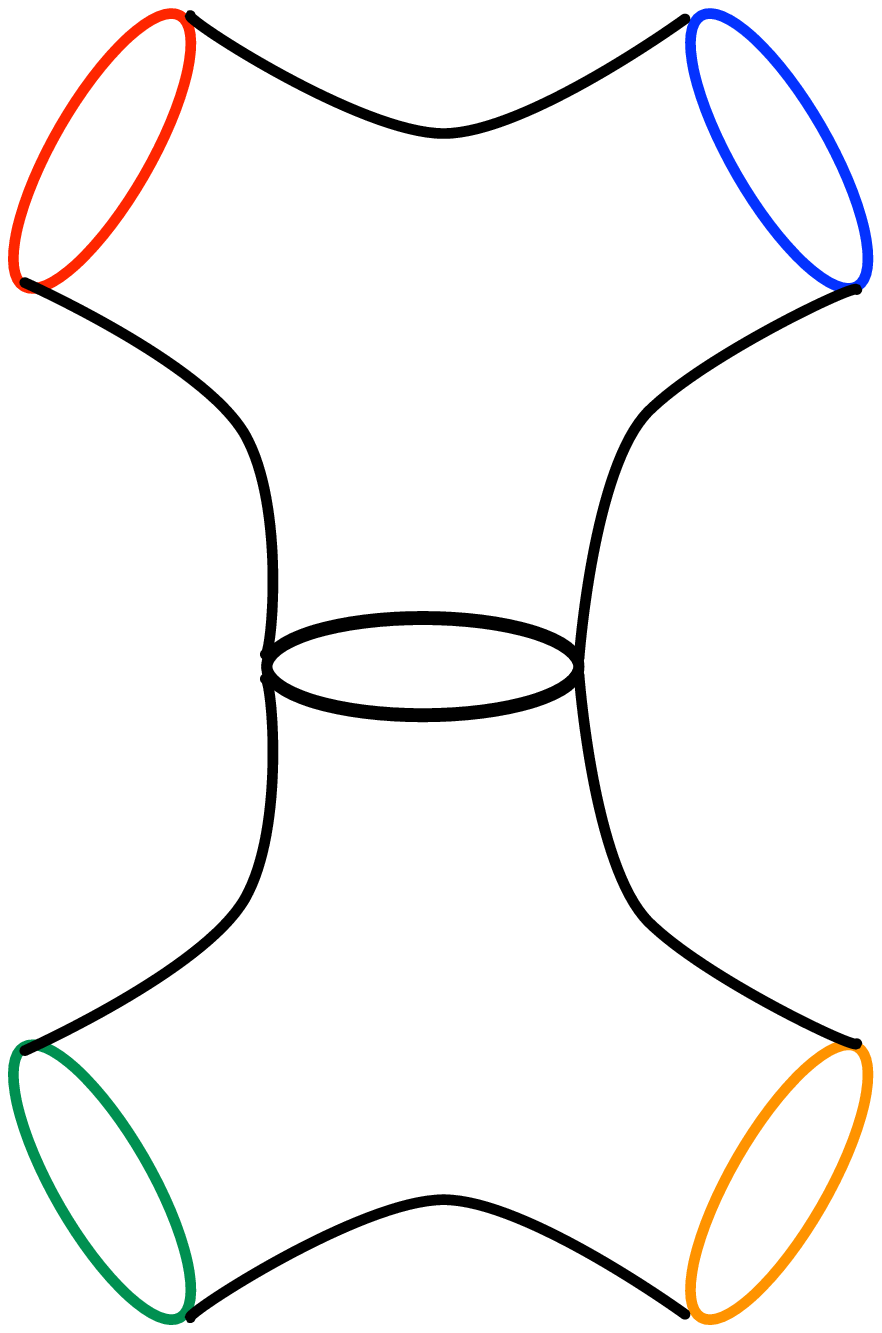}).
\end{equation}
\end{itemize}
We can summarize as follows: there is a one-to-one correspondence between a 2d topological field theory
 $\eta_{G_{\bC}}$ valued in $\HS$ and a pair of holomorphic symplectic varieties $(U_{G_{\bC}},W_{G_{\bC}})$ satisfying the above three properties.
We close with two remarks:
\begin{enumerate}
\item The axioms imply that $W_{G_{\bC}}{}^{n-2}$ has a holomorphic symplectic action of $\fS_{n}\ltimes {G_{\bC}}^{n}$ for all positive $n$. Here we formally take $W_{G_{\bC}}{}^{-1}=U_{G_{\bC}}$ and $W_{G_{\bC}}{}^0=V_{G_{\bC}}=T^*G_\bC$.
\item $\eta_G(\text{torus})$   is the symplectic quotient of $T^*G_{\bC}$ by the
adjoint action of $G_{\bC}$. This shows that the holomorphic varieties are in general singular.

\end{enumerate}

\subsection{Physical data}
In the context of theories of class $S$, physical arguments show that $U_{G_{\bC}}$ is a moduli space of Nahm's equation on a segment.
 These spaces have been studied by Kronheimer \cite{Kronheimer} and Bielawski \cite{Bielawski,Bielawski2}. In particular   $U_{G_{\bC}}$ is given by \begin{equation}
U_{G_{\bC}} = {G_{\bC}}\times S_n \subset {G_{\bC}}\times \fg_\bC \simeq T^*{G_{\bC}}
\end{equation} where $S_n$ is the Slodowy slice at a principal nilpotent element $n$.
Recall that, by definition, the Slodowy slice $S_e\subset \fg_\bC $ at a nilpotent element $e$ is:
 \begin{equation}
S_e =\{  e+v  \in \fg_\bC \ |\  [f,v]=0  \}
\end{equation} where $(e,h,f)$ is an $\sl(2)$ triple containing $e$, i.e. $[e,f]=h$, $[h,e]=2e$ and $[h,f]=-2f$.
Note that the  principal nilpotent element $n$ is unique up to conjugacy so the equivalence class $[U_{G_{\bC}}]$
doesn't depend on the choice of $n$.
For example, for $G_{\bC}=\SL(N,\bC)$, an example of a principal nilpotent element $n$ is a Jordan block of size $N\times N$, and $S_n$ is an affine space of dimension $N-1$.
For example, for $N=4$, we have
\begin{equation}
n=\left(\begin{smallmatrix}
0&1&0&0\\
0&0&1&0\\
0&0&0&1\\
0&0&0&0
\end{smallmatrix}\right)
\qquad\text{and}\qquad
S_n= \Bigl\{ \left(\begin{smallmatrix}
0&1&0&0\\
a&0&1&0\\
b&a&0&1\\
c&b&a&0
\end{smallmatrix}\right) \Bigr\}.
\end{equation}

Now the physical theories of class $S$ predict the existence of a variety $W_{G_{\bC}}$ satisfying the properties above
needed to define a topological field theory $\eta_{G_{\bC}}$. We will give some explicit examples of
$W_{G_{\bC}}$ in the next section but for general ${G_{\bC}}$, $W_{G_{\bC}}$ does not appear to be known, \
and we urge mathematicians to construct it.

We remark that the dimension of $W_{G_{\bC}}$ is easily computed.
The dimension of $U_{G_{\bC}}$ is given by \begin{equation}
\dim_\bC U_{G_{\bC}}=\dim_\bC {G_{\bC}} + \rank {G_{\bC}}.
\end{equation}
Now, since $G_{\bC}$ acts effectively, the    capping axiom implies that
 \begin{equation}
\dim_\bC W_{G_{\bC}}= 3 \dim_\bC {G_{\bC}} - \rank {G_{\bC}},
\end{equation}
It would be interesting to know if there is a unique  holomorphic symplectic
manifold satisfying the above criteria.

\section{Examples}

\subsection{The case $\fg=A_1$}\label{4.1}
For $\fg=A_1$, $W_{G_{\bC}}$ is given by the flat symplectic space $W_{G_{\bC}}=\bC^2\otimes \bC^2\otimes \bC^2$ \cite{Gaiotto:2009we}.
It satisfies all the properties listed in the above
under the natural $\SL(2,\bC)^3$  action, together with
the $\fS_3$ action permuting the three $\bC^2$ factors.
The associativity is somewhat nontrivial.  It turns out that \begin{equation}
\eta_{A_1}(\inc{t-channel}) = (\bC^2\otimes \bC^2\otimes \bC^2) \times (\bC^2\otimes \bC^2\otimes \bC^2) // \SL(2,\bC)
\end{equation} is the Atiyah-Drinfeld-Hitchin-Manin construction of the (closure of the) minimal nilpotent orbit of $\SO(8,\bC)$, or equivalently the framed centered 1-instanton moduli space of $\SO(8)$ gauge fields on $\bR^4$ \cite{Douglas:1996sw}.
$\SO(8,\bC)$ has $\SL(2,\bC)^4$ subgroup, as shown in Fig.~\ref{e}.
The outer automorphism of $\SO(8)$ then provides the action of $\fS_4$, see Sec.~10 of \cite{Seiberg:1994aj}

\subsection{The case $\fg=A_2$}\label{4.2}

\begin{figure}
\[
\includegraphics[scale=.3]{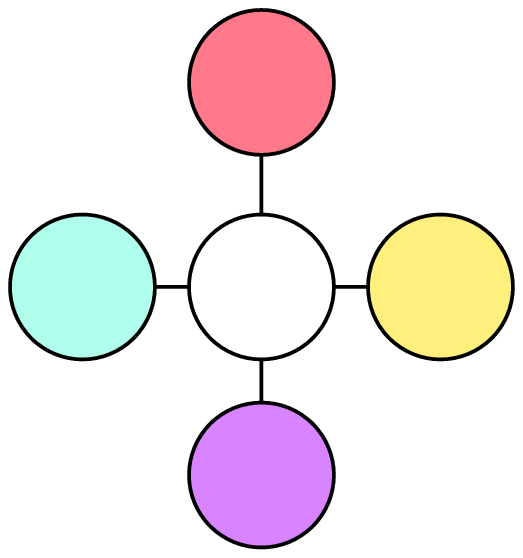}\quad
\includegraphics[scale=.3]{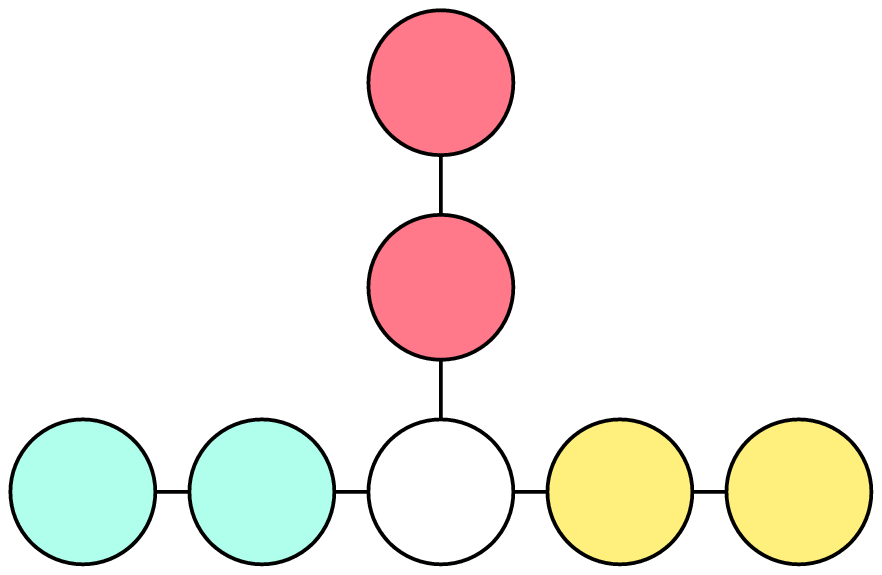}\quad
\includegraphics[scale=.3]{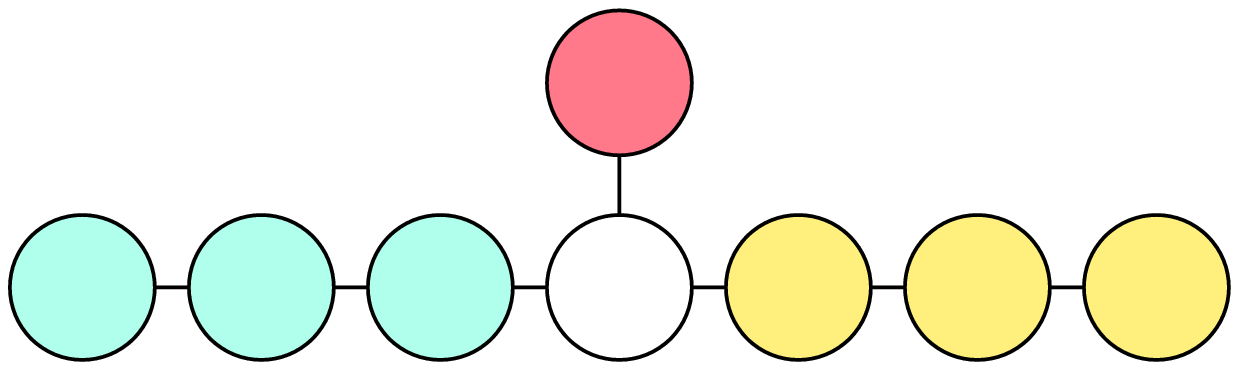}\quad
\includegraphics[scale=.3]{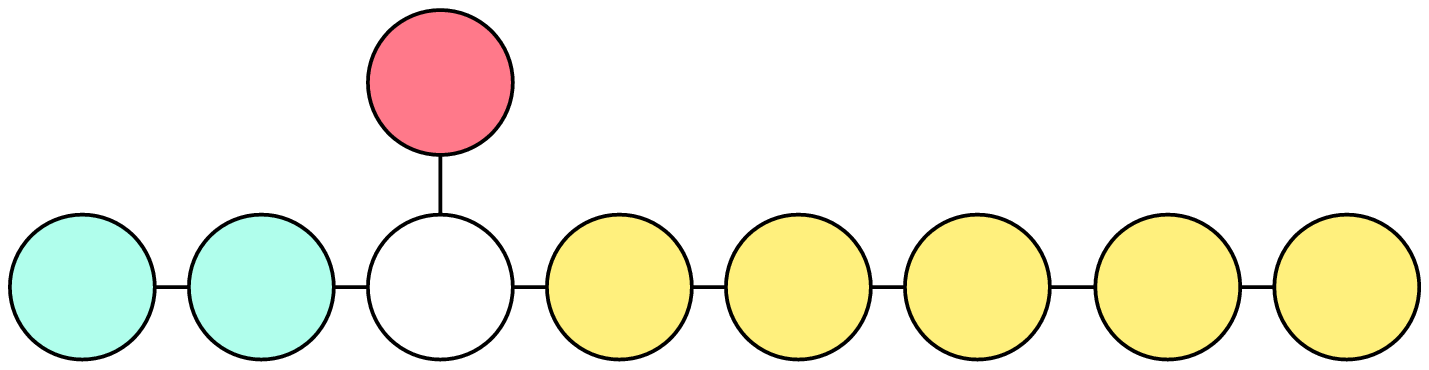}
\]
\caption{Extended Dynkin diagrams of $D_4$, $E_6$, $E_7$ and $E_8$, together with subdiagrams corresponding to $\SL(2)^4$, $\SL(3)^3$, $\SL(4)^2\times \SL(2)$, and $\SL(6)\times \SL(3)\times \SL(2)$ subgroups \label{e}}
\end{figure}

For $\fg=A_2$, $W_{G_{\bC}}$ is believed to be the (closure of the) minimal nilpotent orbit of $E_6(\bC)$ \cite{Argyres:2007cn,Gaiotto:2008nz,Gaiotto:2009we}.
The group $E_6(\bC)$ has a special maximal proper subgroup $\SL(3)^3$, determined from the extended Dynkin diagram shown in Fig.~\ref{e}.
Outer automorphisms of $E_6(\bC)$ provide the action of $\fS_3$ on $W_{G_{\bC}}$ permuting the three $\SL(3)$ actions.
The capping and the associativity have not yet been checked. We believe
 it is a straightforward calculation.

\subsection{A general conjectural property of $W_{G_{\bC}}$}

$W_{G_\bC}$ has moment maps of the $G^3$ action \begin{equation}
\mu_i : W_{G_\bC}\to \fg_\bC^*
\end{equation}  for $i=1,2,3$.
From string theory analysis, it is believed (see e.g. Appendix C of \cite{Benini:2009mz}) that for any element of $P\in \bC[\fg_\bC]^{G_\bC}$ we have \begin{equation}
P (\mu_1)=P(\mu_2)=P(\mu_3)\label{foo}
\end{equation} where we regarded $P$ as a function on $\fg_\bC^*$.
This equality  for $W_{A_1}=\bC^2\otimes\bC^2\otimes\bC^2 $ for the quadratic $P$ reduces to the uniqueness of  Cayley's hyperdeterminant.
This equality for $W_{A_2}$, which is the minimal nilpotent orbit of $E_6(\bC)$,  can be checked using its defining equations due to Joseph \cite{Joseph}.

\subsection{Some more conjectural properties for $\fg=A_n$}

There are a few more results about $W_{A_n}$ deduced from string theory analysis for general $n$.
To describe them, we need to generalize the construction slightly.
We enlarge the morphisms of the source category to be two-dimensional surface with marked points, with additional data at each marked point given by a homomorphism $\rho:\sl(2)\to \fg_\bC$.
Correspondingly, we introduce holomorphic symplectic manifolds \begin{equation}
\eta_{G_{\bC}}(\inc{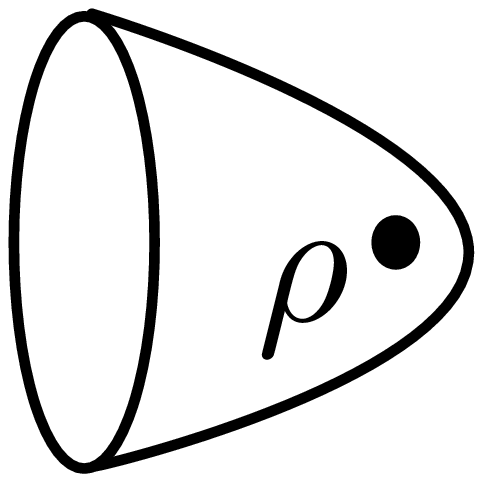})= {G_{\bC}}\times S_{\rho(e)} \subset {G_{\bC}}\times \fg \simeq T^*{G_{\bC}}
\end{equation}
which are also the moduli spaces of Nahm's equation on a segment with appropriate boundary conditions \cite{Bielawski}.
Note that this variety has a Hamiltonian action of ${G_{\bC}}\times Z(\rho)$, where $Z(\rho)$ is the centralizer of $\rho(\SL(2))$ inside ${G_{\bC}}$.
Note also that for $\rho=0$ this manifold is $T^*{G_{\bC}}$ itself.

Then we can associate to a sphere with three marked points a holomorphic symplectic variety using the holomorphic symplectic quotient specified by the following figure: \begin{equation}
\eta_{G_{\bC}}(\inc{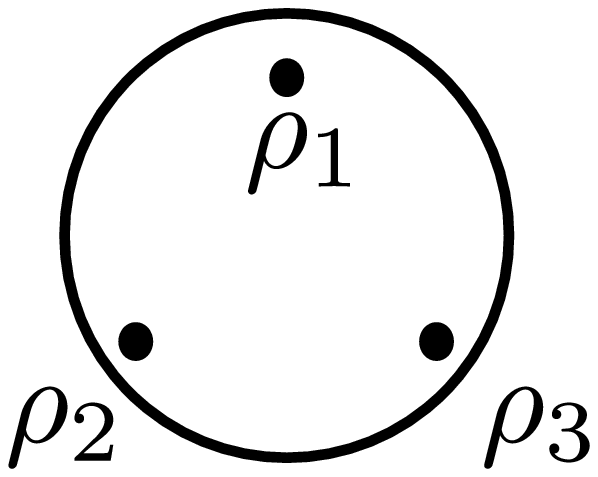})=
\eta_{G_{\bC}}(\inc{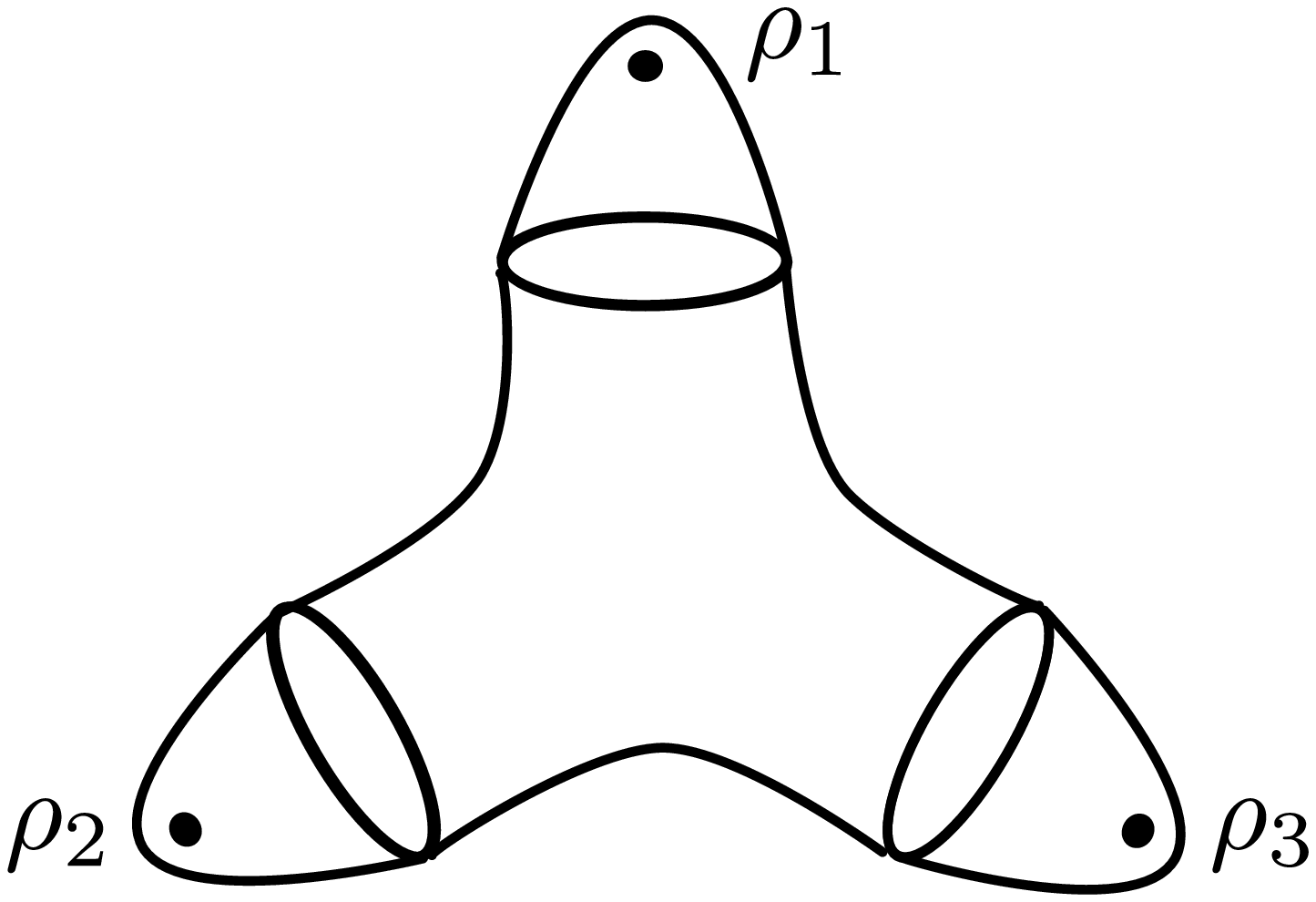}).
\end{equation}
We denote this variety by $\eta_{G_{\bC}}(\rho_1,\rho_2,\rho_3)$.
Similarly, we can define the holomorphic symplectic manifolds $\eta_{G_{\bC}}(\rho_1,\rho_2,\rho_3,\rho_4)$ associated to a sphere with four punctures marked by $\rho_{1,2,3,4}$.

For $G_{\bC}=SL(N,\mathbb{C})$, the homomorphism $\rho:\sl(2)\to \sl(N)$ can be identified with a partition of $N$, which we denote by $[i^{e_i}, j^{e_j}, \ldots]$ for $N=\sum_i e_i i$. In this notation $\rho=0$ corresponds to $[1^N]$.
Now we can start listing the known properties of $\eta_{A_{N-1}}(\rho_1,\rho_2,\rho_3)$.

\subsubsection{Symplectic vector spaces}
With suitable choices of $\rho_{1,2,3}$, we can realize symplectic vector spaces in various representation of $\SL(N)$. For example,
\begin{align}
\eta_{A_{N-1}}([1^N],[N-1,1],[1^N])& =V\otimes V^* \oplus V^*\otimes V, \label{zot}\\
\eta_{A_{N-1}}([1^N],[\lfloor\textstyle\frac{N+1}2\rfloor,\lfloor\textstyle\frac{N}2\rfloor],[\lfloor\textstyle\frac{N}2\rfloor,\lfloor\textstyle\frac{N-1}2\rfloor,1])
&=\wedge^2 V \oplus \wedge^2 V^* \oplus V\otimes \bC^2 \oplus V^*\otimes \bC^2
\end{align} where $V=\bC^N$ \cite{Gaiotto:2009we,Nanopoulos:2009xe,Chacaltana:2010ks}.
More examples for $A_3$, $A_4$ and $D_4$ can be found in \cite{Chacaltana:2010ks,Chacaltana:2011ze}.

From \eqref{zot}, we can deduce that \begin{multline}
\eta_{A_{N-1}}([1^N],\underbrace{[N-1,1],\ldots,[N-1,1]}_k,[1^N]) \\
= \left[\bigoplus_{i=1}^k V_i\otimes V_{i+1}^* \oplus V_i^*\otimes V_{i+1}\right] \bigm/\bigm/  \SL(V_2)\times \SL(V_3)\times \cdots \times \SL(V_k)
\end{multline} where $V_i\simeq  \bC^N$. This is an $\SL$-version of Nakajima's quiver varieties \cite{Nakajima}.

\subsubsection{Instanton moduli spaces}
When the choices of $\rho_{1,2,3}$ are related to the structure of an extended Dynkin diagram, $\eta_{A_n}(\rho_1,\rho_2,\rho_3)$
can sometimes be identified with  instanton moduli spaces \cite{Benini:2009gi,Benini:2010uu}:
\begin{itemize}
\item $\eta_{A_{2k-1}}([k^2],[k^2],[k^2],[k^2])$ is the framed centered $k$-instanton moduli space of $D_4$ on $\bR^4$ of dimension $12k-2$.  Note again the special maximal proper subgroup $\SL(2)^4$ of $D_4$, as shown in Fig.~\ref{e}.
Note that this reduces to the statement on $W_{A_1}\circ W_{A_1}$  in Section \ref{4.2} when $k=1$.
\item $\eta_{A_{3k-1}}([k^3],[k^3],[k^3])$ is the framed centered $k$-instanton moduli space of $E_6$ on $\bR^4$ of dimension $24k-2$. Note that $Z([k^3])=\SL(3)$ and $\SL(3)^3$ is a special maximal proper subgroup of $E_6$.
Note also that this reduces to the statement on $W_{A_2}$ in Section \ref{4.1} when $k=1$.
\item $\eta_{A_{4k-1}}([k^4],[k^4],[2k^2])$ is the framed centered $k$-instanton moduli space of $E_7$ on $\bR^4$ of dimension $36k-2$.
Note that $Z([k^4])=\SL(4)$ and $Z([2k^2])=\SL(2)$.   $\SL(4)^2\times \SL(2)$ is a special maximal proper subgroup of $E_7$ as shown in Fig.~\ref{e}.
\item $\eta_{A_{6k-1}}([k^6],[2k^3],[3k^2])$ is the framed centered $k$-instanton moduli space of $E_8$ on $\bR^4$ of dimension $60k-2$.  Note again the special maximal proper subgroup $\SL(6)\times \SL(3)\times \SL(2)$ of $E_8$, as shown in Fig.~\ref{e}.
\end{itemize}
The analysis of the case $k=1$ goes back to \cite{Banks:1996nj,Minahan:1996fg,Minahan:1996cj,Intriligator:1996ex}.

\section{Functor to the category of hyperk\"ahler manifolds}
Before ending this note, let us briefly discuss why we chose the holomorphic symplectic varieties as the target category.
The Higgs branch of a supersymmetric theory not only comes with a holomorphic symplectic structure, but is equipped with hyperk\"ahler structure. The category $\HK$ of hyperk\"ahler manifolds with triholomorphic action can be defined naturally
by saying that elements of $\Obj(\HK)$ are compact semi-simple groups and that $\Hom(G,G')$ consists of hyperk\"ahler manifolds with triholomorphic action of $G\times G'$.
It seems more natural to take $\HK$ as the target category, but it turns out that $\eta_G$ in the hyperk\"ahler sense is not quite a topological quantum field theory.

The subtlety can be understood by considering $T^*G_\bC \in \Hom(G,G)$, whose hyperk\"ahler structure was originally constructed by Kronheimer using the Nahm equation \cite{Kronheimer}. The hyperk\"ahler metric $g$ can be replaced by $g/a$ where $a\in \bR_+$ is a positive real number without destroying the $G\times G$ invariance or hyperk\"ahler structure; let us denote the resulting rescaled hyperk\"ahler space by $T^*G_\bC{}^a$.  It can be checked that \begin{equation}
T^*G_\bC{}^a \times T^*G_\bC{}^{a'} /\!/\!/ G = T^*G_\bC{}^{a+a'},
\end{equation}
i.e.~the hyperk\"ahler quotient changes the overall factor of the metric.
This fact follows naturally by the fact that $T^*G_\bC{}^a$ is the moduli space of Nahm's equation on a segment of length $a$.
Then, to have an identity in $\Hom(G,G)$, we need to take the $a\to0$ limit of the Riemannian manifold $T^*G_\bC{}^a$, which does not exist in the usual sense.

It seems likely that $\eta_G$ becomes a functor to $\HK$ if  we change the source to be the category without identity of 2-bordisms with the area, so that a morphism is a pair $(C,a)$ where $C$ is an orientable 2-manifold with boundaries and  $a$ is a positive real number which can be thought of as the area of $C$.
The composition of two morphisms then adds the area.
Then, for example, $\eta_{A_1}(W,a)$ is a hyperk\"ahler manifold which is equivalent to $\bC^2\otimes \bC^2\otimes \bC^2$ as a holomorphic symplectic manifold for any $a$; there is a sense in which the $a\to0$ limit of $\eta_{A_1}(W,a)$ is the flat hyperk\"ahler metric on $\bC^2\otimes \bC^2\otimes \bC^2$.
Similarly, $\eta_{A_2}(W,a)$ is equivalent to the minimal nilpotent orbit of $E_6(\bC)$ as the hyperk\"ahler manifold only in the $a\to 0$ limit; at finite $a$, the hyperk\"ahler metric is only invariant under $\SU(3)^3$, not under the full $E_6$. These points need to be studied more carefully.

\section{Further Extensions}

To conclude we would like to mention two further directions in which this
work could be extended. Both extensions appear to us to be nontrivial open
problems.

First, quite generally,  when  the target category has an action of a discrete group $\Gamma$ a
natural generalization of a topological field theory is to the equivariant case, where the
cobordism category is a category of principal $\Gamma$-bundles over a surface, where
$\Gamma$ is a discrete group. The sewing axioms are known in this case
 and can be found in \cite{Moore:2006dw}.
 In our case, we can pick as $\Gamma$ a discrete subgroup of $\Aut(G)$ if $G$ is simply-laced.
Physics predicts the existence of $\Gamma$-equivariant topological field theory
extending the one we have described \cite{Tachikawa:2010vg}.
In particular, the principal $\Gamma$-bundle over $S^1$ with holonomy $x\in\Gamma$
is mapped to $(G^x)^\vee$, where $^\vee$ is the Langlands dual.

Second, it is natural to ask whether the topological field theory
we have described fits nicely into the structure of an
extended topological field theory in the sense described in
\cite{Segal1,Segal2,MR2555928,Freed:2009qp,Kapustin:2010ta}.  Given the relation
of theories of class $S$ to conformal field theory it is natural
to expect a $0$-$1$-$2$-$3$ theory. The extension to level $3$ should
involve Lagrangian subvarieties of the holomorphic symplectic manifolds
and be related to the physics of domain walls in theories of class $S$.

\section*{Acknowledgements}
The authors thank Francesco Benini, Dan Freed, Ron Donagi, Davide Gaiotto, Hiraku Nakajima, Andy Neitzke
and Graeme Segal for discussions. We also thank Daniel Freed and Daniel Roggenkamp for remarks on the draft.
The  work of YT is supported in part by NSF grant PHY-0969448  and by the Marvin L. Goldberger membership through the Institute for Advanced Study.
The  work of YT is also supported in part by World Premier International Research Center Initiative (WPI Initiative),  MEXT, Japan through the Institute for the Physics and Mathematics of the Universe, the University of Tokyo.
The work of GM is supported by the DOE under grant
DE-FG02-96ER40959.

\bibliographystyle{ytphys}
\small\baselineskip=.9\baselineskip
\bibliography{ref}
\end{document}